\RequirePackage{amsmath}

\documentclass[runningheads]{llncs}
\usepackage[T1]{fontenc}
\usepackage{graphicx,verbatim}
\usepackage{booktabs}
\usepackage{multirow}
\usepackage{dsfont}
\usepackage{amsfonts}
\usepackage{amssymb}
\usepackage{subcaption}
\usepackage{xcolor}
\usepackage{color}

\usepackage{xcolor} 
\usepackage[colorlinks=true,
            linkcolor=customorange,
            citecolor=customorange,
            urlcolor=customgreen,
            pdfborder={0 0 0}]
           {hyperref}

\definecolor{customorange}{RGB}{247,120,30}
\definecolor{customgreen}{RGB}{34,139,34} 

\newcommand{\zsem}[0]{z_\text{sem}}

\begin{document}
\title{Contrastive Anatomy-Contrast Disentanglement: A Domain-General MRI Harmonization Method}
\titlerunning{A Domain-General MRI Harmonization Method}

\author{
    Daniel Scholz\inst{1,2,3} \and
    Ayhan Can Erdur\inst{3,4}\and
    Robbie Holland \inst{6} \and
    Viktoria Ehm\inst{2,5} \and
    Jan C. Peeken\inst{4,7,8} \and
    Benedikt Wiestler\inst{1,2,*} \and
    Daniel Rueckert\inst{2,3,*}
}

\authorrunning{D. Scholz et al.}

\institute{%
    Chair for AI for Image-Guided Diagnosis and Therapy, Technical University of Munich (TUM) and TUM University Hospital, Munich, Germany \and
    Munich Center for Machine Learning (MCML), Munich, Germany \and
    Chair for AI in Healthcare and Medicine, Technical University of Munich (TUM) and TUM University Hospital, Munich, Germany \and
    Department of Radiation Oncology, TUM University Hospital, Munich, Germany \and
    Chair for Computer Vision and Artificial Intelligence, Technical University of Munich (TUM), Munich, Germany \and
    Stanford Center for Artificial Intelligence in Medicine and Imaging, Stanford University, Stanford, CA, USA \and
    Deutsches Konsortium für Translationale Krebsforschung (DKTK), Partner Site Munich, Munich, Germany \and
    Institute of Radiation Medicine (IRM), Department of Radiation Sciences (DRS), Helmholtz Center Munich, Munich, Germany \\
    *contributed equally as senior authors \\
    \email{daniel.scholz@mri.tum.de}
}

\maketitle              

\setcounter{footnote}{0}

\begin{abstract}
    Magnetic resonance imaging (MRI) is an invaluable tool for clinical and research applications.
    Yet, variations in scanners and acquisition parameters cause inconsistencies in image contrast, hindering data comparability and reproducibility across datasets and clinical studies.
    Existing scanner harmonization methods, designed to address this challenge face limitations, such as requiring traveling subjects or struggling to generalize to unseen domains.
    We propose a novel approach using a conditioned diffusion autoencoder with a contrastive loss and domain-agnostic contrast augmentation to harmonize MR images across scanners while preserving subject-specific anatomy.
    Our method enables brain MRI synthesis from a single reference image.
    It outperforms baseline techniques, achieving a ${+7}\%$ PSNR improvement on a traveling subjects dataset and +18\% improvement on age regression in unseen scanners.
    Our model provides robust, effective harmonization of brain MRIs to target scanners without requiring fine-tuning.
    This advancement promises to enhance comparability, reproducibility, and generalizability in multi-site and longitudinal clinical studies, ultimately contributing to improved healthcare outcomes.\footnote{The code is publicly available at: \url{https://github.com/daniel-scholz/cacd}.}
\end{abstract}

\section{Introduction}
Magnetic resonance imaging (MRI) is a versatile technique for brain studies, offering multiple tissue contrasts in a single session and enabling non-invasive insights into the brain's structure and function.
However, its flexibility leads to a lack of standardization across imaging studies.
Variations in pulse sequences, acquisition parameters, and scanner hardware can cause undesired contrast variations, particularly in multi-site and longitudinal studies \cite{kusholEffectsMRIScanner2023,hedgesReliabilityStructuralMRI2022}.
These variations create a domain shift problem, introducing underlying bias into the models and limiting the generalizability of quantitative analyses.
Addressing these challenges is crucial for advancing multi-center MRI research and improving diagnostic reliability across healthcare settings.

By using traveling subjects, i.e., scans of the same subject in different scanners serving as ground truth, harmonization can be achieved by learning direct mappings between scanners~\cite{dewey2019deepharmony,wolleb2022diffusion}.
However, traveling subjects require significant effort and hence occur only seldom in clinical reality, limiting the application of such methods to a narrow context.
Many methods alleviate the necessity for traveling subjects through inter-domain cycle consistency~\cite{ozbey2023unsupervised,zhao2019harmonization}.
These methods can usually translate between only two scanners by their pair-wise design, thus lacking the generalizability to harmonize multiple sources into a common space.
Furthermore, the absence of anatomical constraints can lead to unrealistic transformations or even hallucinations.

Recent methods~\cite{zuo2023haca3,cackowski2023imunity} tackle scanner harmonization by disentangling contrast from anatomy, i.e., keeping subject-specific anatomical aspects consistent while learning contrast changes that originate from scanners to generalize to more than two domains.

To this end, Zuo et al.~\cite{zuo2023haca3} rely on paired MRI sequences, such as T1w, T2w, and FLAIR, of the same subject, which are more common in clinical practice, to learn anatomical consistency.
Models that can achieve disentanglement from only a single sequence also exist \cite{cackowski2023imunity}.
However, due to the lack of domain generalization, these models are limited to synthesizing images similar to their training domain.
In particular, introducing new domains unseen during training often fails but is a clinical reality.

In our work, we develop an anatomy-consistent approach that harmonizes between an arbitrary number of unseen scanners outside the training domain.
Additionally, the proposed method overcomes the reliance on paired subjects or multiple MRI sequences per subject.

We summarize our contributions as follows:
\begin{enumerate}
    \item We present a domain-general scanner harmonization algorithm capable of \textbf{harmonizing arbitrary unseen scanners} in brain MRI by controlling contrast and maintaining anatomy.
    \item To this end, we introduce \textbf{a novel anatomy-contrast-disentanglement module}, including an elegant loss formulation with contrastive losses and a domain-agnostic contrast augmentation suite.
    \item We \textbf{advance MRI scanner harmonization in unseen scanners} in terms of traveling subjects harmonization and age regression.
\end{enumerate}

\section{Related Works}

Numerous methods have been proposed for MRI harmonization, which are discussed in detail in these two recent reviews~\cite{gebreCrossScannerHarmonization2023a,huImageHarmonizationReview2023}, so that we focus on discussing approaches most relevant to our work.
\paragraph{Learning-based Methods}
Many existing deep learning models for scanner harmonization require traveling subjects~\cite{dewey2019deepharmony,wolleb2022diffusion}.
Their reliance on full supervision, i.e., datasets with images of the same patient from multiple sites, limits their applicability drastically.

In contrast, the unsupervised CycleGAN method~\cite{zhu2017unpaired} uses a cycle consistency loss to learn mappings between two domains without paired examples.

While adaptations of CycleGAN have been explored for MRI harmonization~\cite{ozbey2023unsupervised,zhao2019harmonization}, this approach can naturally only handle two sites, and anatomical consistency during harmonization is not ensured.
\paragraph{Latent-Space Disentanglement}
Representation disentanglement has been a significant focus of the computer vision community, e.g. for style transfer~\cite{gatysImageStyleTransfer2016,kotovenkoContentStyleDisentanglement2019} or medical image segmentation~\cite{gu2023cddsa,huDevilChannelsContrastive2023}.
Some works accomplish scanner harmonization with disentangled latent spaces to maintain anatomical consistency while learning intensity-level translation \cite{cackowski2023imunity,zuo2023haca3}. HACA3~\cite{zuo2023haca3} extract the anatomy embedding through shape constraints from paired MRI sequences while the contrast is learned as the remaining information.
To overcome the need for paired MRI sequences, \cite{cackowski2023imunity} introduce Gamma augmentations to create different contrasts.

This approach, however, is limited to the contrast differences covered by Gamma augmentations, which do not account for the entire variation in real-world data, and relies on patches for training.
These limitations highlight the need for a framework capable of processing full volumes while leveraging a more generalizable image synthesis and latent-space disentanglement.

\section{Method}

\subsection{Diffusion Autoencoder}
The foundation of our work is the diffusion autoencoder (DiffAE)~\cite{preechakulDiffusionAutoencodersMeaningful2022}.

It consists of a conditional DDIM \cite{songDenoisingDiffusionImplicit2022} and a separate semantic encoder $E(x)$ that projects the input image $x$ to a latent feature vector $\zsem$, which is then used to condition the DDIM.
The diffusion autoencoder architecture allows control of the image editing process by altering the semantic feature vector to contain the semantic information desired in the edited image.
However, the feature space needs to be disentangled to make specific changes that leave other imaging features constant.

\begin{figure}[ht]
    \centering
    \includegraphics[width=\linewidth]{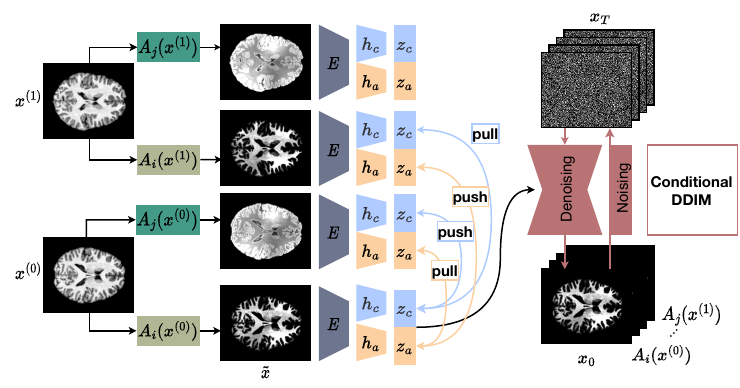}
    \caption{\textbf{Our novel anatomy-content-disentanglement module}.
        Two images from different subjects and arbitrary (possibly even the same) scanners $x^{(0)}$ and $x^{(1)}$ are modified through two augmentations $A_j(x)$ and $A_i(x)$, yielding a view $\tilde{x}$ each.
        An anatomy-contrast disentanglement is learned through deliberate pushing and pulling (see contrastive loss, Eq.~\ref{eq:supervised-contrastive-loss}) of the anatomy $z_a=h_a(E(\tilde{x}))$ and contrast $z_c = h_c(E(\tilde{x}))$ embeddings (Section \ref{DisEntMod}).
        Note that in this example, push and pull are shown only for a single view - in practice, all views are considered.
        The resulting feature vectors $z_a$ and $z_c$ are used to condition the DDIM to synthesize  $\tilde{x}$ by denoising the noise map $x_T$ to the input view denoted as $x_0 \hat{=} \tilde{x}$.
    }
    \label{fig:pipeline}
\end{figure}

\subsection{Disentanglement Module} \label{DisEntMod}

In this work, we aim to disentangle anatomy and contrast to enable altering intensity while maintaining the anatomy in the image~(Figure~\ref{fig:pipeline}).
To this end, we split the feature vector $\zsem \in \mathbb{R}^d$ into two feature vectors $z_a\in \mathbb{R}^{d_a}$ and $z_c\in \mathbb{R}^{d_c}$ for anatomy and contrast, respectively.
To achieve this, we add two separate network heads such that $h_a\left(E\left(x\right)\right) = z_a$ and $h_c\left(E\left(x\right)\right) = z_c$.

To disentangle the learned anatomical and contrast vectors, we introduce a novel loss formulation based on a contrastive loss~\cite{chenSimpleFrameworkContrastive2020b,khosla2020supervised}.
First, we sample a set of $N_\text{augs}$ contrast augmentations $\{A_j(x)\}_{j\in \{0, \dots, N_\text{augs}\}}$, which model a variety of intensities, but, crucially, preserve the anatomy (Section \ref{sec:view-generation}).
We then apply each augmentation to each image $x^{(i)}$ in a batch of images such that we end up with augmented versions of the input images, termed views, $\tilde{x}^{(i)}$ that have greatly varying contrasts but maintain the original anatomy of $x^{(i)}$.
The total number of views generated is $B\cdot N_\text{augs}$, where $B$ is the original number of input images.
Finally, each augmented view $A_j(x^{(i)})$ is projected into an anatomy $z_a^{(i)}$ and contrast vector $z_c^{(i)}$ using the encoder and respective network heads.
For simplicity, we denote $z_a^{(i)}=z_a$ and $z_c^{(i)}=z_c$.

Next, we utilize the supervised contrastive loss~\cite{khosla2020supervised}, an established modificaiton of SimCLR~\cite{chenSimpleFrameworkContrastive2020b} (Eq. \ref{eq:supervised-contrastive-loss}).
This formulation includes the temperature-scaled cosine similarity $\text{sim}$, with temperature $\tau$.
This loss is minimized by requiring the feature vectors of positive pairs in $\mathcal{P}^{(i)}$ of a view $z^{(i)} \in \mathbb{R}^{d}$ to be similar (\textit{pull}) and negative pairs $\mathcal{N}^{(i)}$ to be dissimilar (\textit{push}):

\begin{equation}
    \ell(z^{(i)}, \mathcal{P}^{(i)},\mathcal{N}^{(i)}) = -\log\frac{\sum_{k\in \mathcal{P}^{(i)}}\exp\left(\text{sim}\left(z^{(i)},z^{(k)}\right)/\tau\right)}{\sum_{k\in (\mathcal{P}^{(i)}\bigcup \mathcal{N}^{(i)})}\exp\left(\text{sim}\left(z^{(i)},z^{(k)}\right)/\tau\right)}
    \label{eq:supervised-contrastive-loss}
\end{equation}

We enforce anatomy-contrast disentanglement by carefully defining positive and negative pairs.

To this end, we choose the positive pairs to be all pairs of views originating from the \emph{same} subject but were transformed with \emph{different} augmentations.
The index set of these positive pair candidates for the $i$-th view in a batch is denoted as $\mathcal{P}_a^{(i)}$.
The negative pairs are built from all views originating from a \emph{different} subject that were transformed with the \emph{same} augmentations.
Correspondingly, the index set of negative view candidates for the $i$-th image in a batch is denoted as $\mathcal{N}_a^{(i)}$.
This design forces the network to find similar representations for the intensity-augmented views, leading to only anatomy information encoded in $z_a$ and, hence, anatomy-contrast disentanglement.

To force the model to extract only contrast information into $z_c$

the definition of positive and negative pairs is ``inverted''.
All positive pairs, denoted as $\mathcal{P}_c^{(i)}$, are given by the views originating from the same augmentation but different subjects, i.e., the shared feature between these views is the contrast.
On the contrary, we consider the views of the same subject that were transformed with different augmentations to be negative, denoted as $\mathcal{N}_c^{(i)}$.
Consequently, the differentiating feature between the negative views is also the contrast.
Overall, the encoder $E(x)$ and the heads $h_a$ and $h_c$ need to learn disentangled representations to minimize the overall loss per augmented view at index $i$:
\begin{equation}
    \mathcal{L}_\text{total}^{(i)} = \mathcal{L}_{\text{DDIM}}^{(i)} + \lambda_{c} \ell(z_c^{(i)}, \mathcal{P}_c^{(i)}, \mathcal{N}_c^{(i)}) + \lambda_a  \ell(z_a^{(i)}, \mathcal{P}_a^{(i)}, \mathcal{N}_a^{(i)})
    \label{eq:total-loss}
\end{equation}
with $\ell(z^{(i)})$  the loss defined in Eq.~\ref{eq:supervised-contrastive-loss}.
$\mathcal{L}_{\text{DDIM}}^{(i)}$ is the standard MSE loss to train the DDIM model~\cite{songDenoisingDiffusionImplicit2022}.
In practice, we set $\lambda_c = \lambda_a = 0.5$.

During inference, we first estimate a target contrast vector $\Bar{z}_c$ by projecting and averaging a set of target images to a contrast vector.
The target anatomy vectors are discarded.
The source image is projected to the latent space where its contrast vector $z_c$ is swapped with $\Bar{z}_c$.
The new feature vectors $\Bar{z}_c,z_a$, i.e., the target contrast and source anatomy, serve as a condition to the diffusion model.

\subsection{Augmentation Strategy}\label{sec:view-generation}
Our augmentation strategy is designed to cover all kinds of scanner variations while preserving the subject's anatomy.
For our augmentation set, we gather augmentation schemes from different previous works, namely, the global intensity non-linear augmentation (GIN)~\cite{ouyangCausalityInspiredSingleSourceDomain2023} and
the Synth\textit{X}-framework~\cite{iglesiasJointSuperresolutionSynthesis2021}.

We apply Gamma augmentations with $\gamma=e^u, u \sim \mathcal{U}(0,0.5)$, and randomly generated biasfield corruptions
as in the Synth\textit{X}-framework.
However, we replace the synthetic contrasts generated in Synth\textit{X} by applying the GIN augmentation~\cite{ouyangCausalityInspiredSingleSourceDomain2023} to real samples from the training dataset, making our method agnostic to the anatomical region.

The augmented image is normalized to the $[{-1},1]$-range and linearly interpolated with the original input image with a uniformly sampled interpolation factor $\alpha$.

To counteract the blurring that occurs with GIN and deeper network sizes, we linearly upsample the images to obtain sharper images that are down-sampled back to the original resolution after augmentation.
Examples are shown in Figure~\ref{fig:pipeline}.
Importantly, the augmented images also serve as targets for the diffusion model, enabling our model to synthesize various intensities.

\subsection{Implementation Details}
We parametrize the latent space heads $h_{\{c, a\}}$ as linear layers.
The latent space dimensions for both $z_a$ and $z_c$ is $d_a = d_c = 256$.
We use an Adam optimizer with a learning rate of $1\mathrm{e}{-4}$.
We train the model for $2.5$ million steps on a single NVIDIA A40 GPU.
Our total number of views is 81, derived from $N_\text{augs}=9$  and $B = 9$ subjects.
The GIN network~\cite{ouyangCausalityInspiredSingleSourceDomain2023} consists of two hidden layers with two hidden channels, a kernel size of 3, and bilinear upsampling to 2048 pixels.

\section{Results}
\subsection{Datasets}
We use three different datasets for training and evaluation of our models: OASIS3~\cite{lamontagne2019oasis}, IXI~\cite{IXIDatasetBrain}, and OpenNeuro(ON)Harmony~\cite{warringtonONHarmonyResourceDevelopment2023}.
The OASIS3 dataset includes 3231 T1w MR Scans from 1047 subjects from various scanners of multiple vendors and field strengths (for more details, see~\cite{lamontagne2019oasis}).
Next, the IXI dataset consists of 580 healthy subjects from three sites: Hammersmith Hospital (HH, Philips 3T), Guy’s Hospital (Philips 1.5T), and the Institute of Psychiatry (IOP, GE 1.5T).
The ONHarmony dataset (Phase A, Version 1.0.2) consists of 10 healthy subjects who have been scanned at least once in six 3T MR scanners from the major vendors Siemens, Philips, and GE, yielding 80 T1w MRIs.
We use 80\% of IXI-\{Guy's, HH\}, and the full OASIS3 dataset to train our model.
The remaining IXI data, i.e., 20\% of IXI-\{Guy's, HH\} and the full IXI-IOP, are used to evaluate our model and compare it to other scanner harmonization methods.
The ONHarmony dataset is used to evaluate traveling subjects, which poses a particularly hard challenge since the scanners are unseen to all methods.

To standardize MRI data, we affinely register all scans to the MNI152 atlas at $1\times1\times1$mm$^3$ resolution using the niftyreg library.
We then skull-strip the scans using HD-BET and apply N4 bias field correction.
Lastly, the images are normalized to the $[{-1},1]$-range in the brainmask.
To save computational resources, we randomly sample axial slices from the middle section (middle slice $\pm 10$) of the scan for training and use the axial middle slice for evaluation.
Our method, however, is not reliant on patches and can thus be readily applied to 3d volumes when the resources are available.

\subsection{Traveling Subjects}
\begin{figure}[ht]
    \begin{subfigure}[t]{0.35\linewidth}
        \centering
        \includegraphics[width=\linewidth]{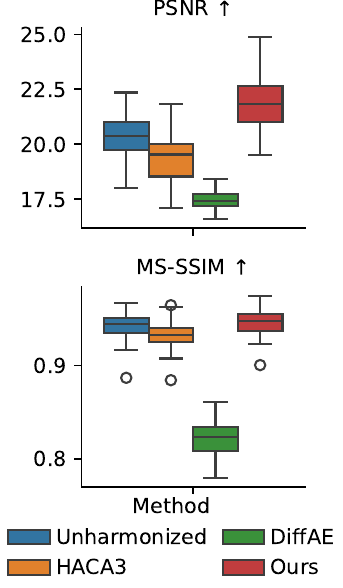}
        \caption{}
        \label{fig:paired-eval}
    \end{subfigure}\hfill
    \begin{subfigure}[t]{0.62\linewidth}
        \centering
        \includegraphics[width=\linewidth]{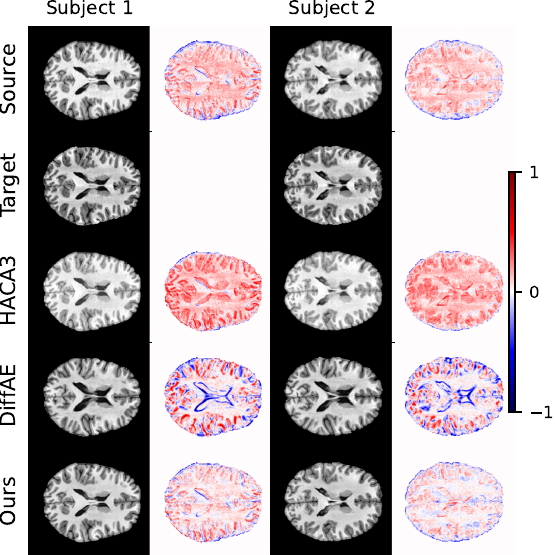}
        \caption{}
        \label{fig:vis-paired-eval}
    \end{subfigure}
    \caption{ \textbf{(a)}~\textbf{Traveling subjects evaluation.}
        Our method outperforms the other methods in both PSNR and MS-SSIM.
        \textbf{(b)}~\textbf{Visual comparison.} The difference maps between the harmonized and target images visualize contrast adjustment while anatomy is maintained (Best viewed magnified on screen).
    }
\end{figure}

To evaluate the efficacy of scanner harmonization we need ground truth data, which requires a subject to be scanned in multiple scanners within a short time period.
To this end, we use paired scans from traveling subjects in the ONHarmony dataset~\cite{warringtonONHarmonyResourceDevelopment2023} and translate the scans from all scanners to the GE scanner present in the dataset, as this is visually the most distinct.
We compare our method to the current state-of-the-art unsupervised deep learning scanner harmonization model HACA3~\cite{zuo2023haca3} with the provided pre-trained weights as well as the standard DiffAE~\cite{preechakulDiffusionAutoencodersMeaningful2022} without disentanglement.
Our quantitative results in Figure~\ref{fig:paired-eval} demonstrate a significant improvement ($p<0.01$,
Wilcoxon signed rank test
) over all baselines in all metrics, especially in PSNR.
The qualitative results in Figure~\ref{fig:vis-paired-eval} support these findings visually.
Our difference maps' lower intensity  indicates a smaller error than the other methods.
Importantly, we also observe anatomically faithful reconstructions and well-adjusted contrasts in the harmonized images regardless of the source scanner or subject due to the large variety of intensities our model has seen during training.
HACA3 reconstructs the anatomy faithfully but fails to capture the contrast correctly due to the unseen scanners confronting the model with unseen intensities.
The standard DiffAE manages to match the intensity profile but significantly changes the anatomy of the subject due to the entanglement of anatomy and contrast by changing ventricle sizes and gyri and sulci shapes.

\subsection{Scanner Classification and Age Regression}

\begin{table}[ht]
    \centering
    \caption{\textbf{Scanner classification and age regression results.}
        We show the effectiveness of our method in harmonizing images by (i) fooling a scanner classifier (\textit{scanner bias}), (ii) achieving high \textit{target fidelity} by translating images to a new domain, and (iii) improving \textit{age regression} performance on an unseen scanner (IOP).
        Bold values indicate the best performance in each column.
    }
    \begin{tabular}{@{}lccccccc}
        \toprule
        \multirow{3}{*}{Method} & \multicolumn{3}{c}{Scanner Bias}      & \multicolumn{3}{c}{Target Fidelity}  & \multicolumn{1}{c}{Age Regression}                                                                              \\ \cmidrule(lr){2-4}\cmidrule(lr){5-7}\cmidrule(lr){8-8}
                                & \multicolumn{3}{c}{F1 ($\downarrow$)} & \multicolumn{3}{c}{Acc ($\uparrow$)} & \multicolumn{1}{c}{$R^2$ ($\uparrow$)}                                                                          \\ \cmidrule(lr){2-4}\cmidrule(lr){5-7}\cmidrule(lr){8-8}
        Target Scanner          & Guy's                                 & HH                                   & IOP                                    & Guy's        & HH           & IOP            & IOP $\rightarrow$ Guy's \\ \midrule
        Unharmonized            & 0.593                                 & 0.568                                & 0.543                                  & 0.031        & 0.222        & 0.021          & 0.58                    \\
        CycleGAN                & \textbf{0.0}                          & 0.4                                  & n/a                                    & \textbf{1.0} & 0.0          & n/a            & -6.53                   \\
        HACA3                   & 0.226                                 & 0.688                                & 0.451                                  & 0.469        & 0.0          & 0.383          & 0.57                    \\
        \textbf{Ours}           & 0.187                                 & \textbf{0.111}                       & \textbf{0.167}                         & 0.75         & \textbf{0.8} & \textbf{0.681} & \textbf{0.69}           \\ \bottomrule
    \end{tabular}
    \label{tab:scanner_bias}
\end{table}

To evaluate the effectiveness of various harmonization methods in mitigating domain shift, we train a radiomics-based classifier~\cite{vangriethuysenComputationalRadiomicsSystem2017} (pyradiomics 3.1.1) on three subsets of the IXI dataset: Guy's, HH, and IOP.
When images are perfectly harmonized to a specific target scanner, the classifier should classify every sample as stemming from this target scanner.
We thus harmonize all images to a specific scanner (column head in Table~\ref{tab:scanner_bias}) and evaluate two scenarios: In the \textit{scanner bias} scenario, we compare the classifier's output to the original scanner labels aiming at minimizing the F1 score ($\downarrow$), contrary to typical objectives.
Conversely, in the \textit{target fidelity} scenario, we compare the classifier's output to the label of the target scanner itself expecting maximal classification accuracy ($\uparrow$).
We compare our method against CycleGAN~\cite{zhu2017unpaired}, trained on Guy's and HH, and HACA3~\cite{zuo2023haca3}.
Our model consistently works robustly for all three scanners, even when choosing IOP as the target - a scanner not seen during training - and outperforms HACA3 across all comparisons.
CycleGAN performs well for the Guy's scanner but fails for HH and is not applicable for IOP, highlighting its limitations.

Finally, as a downstream task, we train an age regressor on MRIs from Guy's and HH, and evaluate it in IOP, either unharmonized or following harmonization to Guy's domain space (Table~\ref{tab:scanner_bias}, "Age Regression").
Our method outperforms all comparisons methods in terms of $R^2$ ($\uparrow$) and significantly improves age regression in unseen scanners.
Overall, our method shows the best domain generalization to unseen scanners, while not needing to train on target domain data.

\section{Discussion and Conclusion}
This work presents a novel method for scanner harmonization of brain MRI.
Our approach builds upon the diffusion autoencoder~\cite{preechakulDiffusionAutoencodersMeaningful2022} and introduces a novel contrastive anatomy-contrast-disentanglement module, including a domain-agnostic intensity augmentation suite. Importantly, our approach does not require paired data, either in the form of multiple scans of the same subject or multiple sequences, underscoring its broad applicability and overcoming a key limitation of earlier approaches~\cite{dewey2019deepharmony,wolleb2022diffusion}.
Currently, our approach uses 2D slices of the volumes due to the heavy computational load of diffusion models.
However, only the synthesis part of our model is challenging to upgrade to 3D, as we do not rely on comparisons between slices, as in~\cite{zuo2022disentangling,cackowski2023imunity}.
A simple remedy to this problem is to use latent or wavelet diffusion models~\cite{rombach2022high,friedrich2024wdm}.
In conclusion, we believe that our work addresses challenges crucial for advancing multi-center MRI research and diagnostic healthcare through generalizable scanner harmonization in the future.

\begin{credits}
    \subsubsection{\ackname}
    This study was supported by the DFG, grant \#504320104.

    \subsubsection{\discintname}
    The authors have no competing interests to declare that are
    relevant to the content of this article

\end{credits}

%
%
%
\bibliographystyle{splncs04}
\bibliography{references.bib}
\end{document}